\documentclass[runningheads,a4paper,oribibl]{llncs}

\usepackage[british]{babel}

\usepackage[%
rm={oldstyle=false,proportional=true},%
sf={oldstyle=false,proportional=true},%
tt={oldstyle=false,proportional=true,variable=true},%
qt=false%
]{cfr-lm}
\usepackage[utf8]{inputenc} 
\usepackage[T1]{fontenc}
\usepackage{microtype}

\usepackage{url}
\makeatletter
\g@addto@macro{\UrlBreaks}{\UrlOrds}
\makeatother

\usepackage{xcolor}

\usepackage[
bookmarks=false,
breaklinks=true,
colorlinks=true,
linkcolor=black,
citecolor=black,
urlcolor=black,
pdfpagelayout=SinglePage,
pdfstartview=Fit
]{hyperref}
\usepackage[all]{hypcap}

\usepackage[%
  sorting=none,
  backend=biber,
  style=lncs,
]{biblatex}
\usepackage{amsmath,amssymb}
\usepackage{enumitem}
\setlist[description]{style=standard}
\usepackage{calc}

\usepackage{tikz}
\usepackage{pgfplots}
\usepackage{pgfplotstable}
\usepgfplotslibrary{statistics}
\pgfplotsset{%
  compat=1.12,
}

\usepackage{tabularx}

\usepackage{graphicx}
\usepackage{subcaption}
\def\tightlist{}
\usepackage{longtable}
\usepackage{booktabs}

\usepackage{siunitx}

\usepackage{csquotes}

\newcommand{\landau}[1]{\ensuremath{\mathcal{O}\left(#1\right)}}

\DeclareSourcemap{
 \maps[datatype=bibtex]{
   \map{
     \step[fieldsource=doi,
       match=\regexp{\{\\_\}},
       replace={2}]
     \step[fieldsource=url,
       match=\regexp{\{\\_\}},
       replace={_}]
   }
 }
}

\DeclareTextCommandDefault{\nobreakspace}{\leavevmode\nobreak\ }
\newcommand\cS{{\mathbb{S}}}

\newcommand\cSs{{\widehat{\mathbb{S}}}}

\usetikzlibrary{backgrounds,calc,fit,decorations.pathreplacing}
\usetikzlibrary{positioning,shapes,shadows,arrows,decorations.pathmorphing}

\begin{document}

\title{Distributed Random Process for a\\
large-scale Peer-to-Peer Lottery}

\author{Stéphane Grumbach \and Robert Riemann}
\institute{Inria \newline \email{\{stephane.grumbach,robert.riemann\}@inria.fr}}

%

\maketitle

\begin{abstract}
Most online lotteries today fail to ensure the verifiability of the
random process and rely on a trusted third party. This issue has
received little attention since the emergence of distributed protocols
like Bitcoin that demonstrated the potential of protocols with no
trusted third party. We argue that the security requirements of online
lotteries are similar to those of online voting, and propose a novel
distributed online lottery protocol that applies techniques developed
for voting applications to an existing lottery protocol. As a result,
the protocol is scalable, provides efficient verification of the random
process and does not rely on a trusted third party nor on assumptions of
bounded computational resources. An early prototype confirms the
feasibility of our approach.
\end{abstract}


\section{Introduction}\label{sec:introduction}

Lottery is a multi-billion dollar industry \autocite{isidore2015}. In
general, players buy lottery tickets from an authority. Using a random
process, e.g.~the drawing of lots, the winning tickets are determined
and the corresponding ticket owners receive a reward.

In some lotteries, the reward may be considerable, and so is the
incentive to cheat. The potential of fraud gained attention due to the
\emph{Hot Lotto} fraud scandal. In 2015, the former security director of
the Multi-State Lottery Association in the US was convicted of rigging a
14.3 million USD drawing by the unauthorised deployment of a
self-destructing malware manipulating the random
process~\autocite{lottofraud2015}.

In order to ensure fair play and ultimately the trust of players,
lotteries are subject to strict legal regulations and employ a technical
procedure, the \emph{lottery protocol}, to prevent fraud and convince
players of the correctness. Ideally, players should not be required to
trust the authority. Verifiable lottery protocols provide therefore
evidence of the correctness of the random process.

In a simple paper-based lottery protocol, tickets are randomly drawn
under public supervision of all players from an urn with all sold
tickets to determine the winners. Afterwards, all tickets left over in
the urn are also drawn to confirm their presence and convince the losers
of the correctness. Without public supervision, the random process can
be repeated until by chance a predefined result occurred. Further, the
process can be replaced entirely by a deterministic process.

In practice, the public supervision limits the number of lottery players
and is further very inconvenient, because players are required to
respect the time and locality of the drawing procedure. With the advent
of public broadcasting channels, first newspapers, then radio and
television broadcasting, protocols were employed that replaced the
public supervision by a public announcement. Only few players and
notaries verify here the correctness of the random process. In
consequence, the majority of non-present players are required to trust
the few present individuals for the sake of scalability. With the
increasing availability of phones and later the internet, protocols have
been adapted to allow also the remote purchase of lottery tickets,
e.g.~from home or a retail store.

The technical evolution lead to a gradual change of how people play
lottery, but in many cases, the drawing procedure has not been adapted
and resembles more a legacy that prevails for nostalgic reasons than to
provide security. In the simple paper-based lottery protocol, the chain
of custody establishes trust. All operations may be inspected by
eye-sight.

This technique cannot be directly adapted for an online lottery. Thus,
most verifiable online lottery
protocols~\autocites{goldschlag1998}{zhou2001}{chow2005} rely on a
concept based on two elements. All players can actively contribute to
the random process. Nobody can compute the random process result or its
estimation as long as it is possible to contribute or buy new tickets.
The latter is required to prevent educated contributions to circumvent
the uniform distribution of the random process. It is the lottery
protocol that must ensure the order in time of the contribution and the
actual determination of winners.

A protocol consisting of equipotent players contributing each to the
randomness of the publicly verifiable random process is promising for
its similarity with the simple paper-based lottery protocol. Again, all
players participate in the execution and supervision of the random
process. The feasibility to construct such protocols with no trusted
third parties has been demonstrated by the crypto-currency
Bitcoin~\autocite{bitcoin08} that is a distributed protocol for remote
financial transactions, while previously online banking based on trusted
financial authorities, the bank institutes and central banks, has been
without alternative~\autocite{perezmarco2016}. Lottery protocols based
on Bitcoin have been already considered \autocite{wesolowski2016},
c.f.~Section~\ref{sec:related-work}.

Although different, the security requirements of lotteries share common
concerns with those of voting systems~\autocite{lambrinoudakis2003}.
Both lottery and voting protocols have to assure trust in an environment
of mutual distrust among players, respectively voters, and the
potentially biased authorities. The literature on voting protocols and
online voting protocols is extensive and comprises flexible protocols
that may be adapted to different voting systems beyond the general case
of majority voting. Of particular interest for a lottery are online
voting protocols that allow for a random choice. Already the paper-based
voting common for general elections provides a solution to improve the
scalability of the simple paper-based lottery protocol: the introduction
of multiple offices run in parallel. We focus on online voting protocols
that do not rely on trusted parties and aim to provide security
properties that we adopt for lottery applications as follows:

\setlist[description]{style=standard}

\begin{description}
\tightlist
\item[Correctness of the random process]
All numbers are equally likely to win. Nobody can predict the random
process better than guessing.
\item[Verifiability of the random process]
Players can be convinced that the random process has not been
manipulated.
\item[Privacy of the player]
Players do not learn the identity of other players to prevent
blackmailing or begging.
\item[Eligibility of the ticket]
Tickets cannot be forged. Especially, it is impossible to create a
winning ticket after the outcome of the random process is known.
\item[Confidentiality of the number]
Numbers are confidential to ensure fairness. Tickets of other players
cannot be copied to reduce their potential reward.
\item[Completeness of the reward]
Players can verify the number of sold tickets that may determine the
reward.
\end{description}

\noindent
Our contribution is a novel protocol for verifiable large-scale online
lotteries with no trusted authority to carry out the random process, for
which we use concepts originating from online voting.

The paper is organized as follows. In the next section, we review
related work addressing both lottery and voting protocols. Our protocol
is based on an existing online lottery protocol and the distributed hash
table Kademlia that are presented in Section~\ref{sec:preliminaries}.
Then, we introduce our protocol in Section~\ref{sec:distributed-lottery}
and discuss its properties in Section~\ref{sec:evaluation}.

\section{Related Work}\label{sec:related-work}

Different protocols have been proposed that allow players to contribute
to the publicly verifiable random process and take measures to prevent
early estimations of the result while it is still possible to
contribute.

A trivial solution in the context of secure parameters for cryptography
is recalled in~\autocite{wesolowski2015}. In a first round, all players
choose secretly a number and publish on a public broadcasting channel a
commitment on their number, e.g.~using a hash. In a second round, all
secret numbers are revealed and verified using the commitment. Finally,
all values are concatenated using the XOR operation to form the result.
The protocol owes its correctness due to the clear separation in two
rounds of player's contributions. However, the authors stress that the
protocol is neither robust nor scalable. A termination is not possible
if one player does not reveal its secret number and for the
verification, all players have to run as many XOR operations and send as
many messages as there are players.

Subsequently, a random process protocol with only one round is proposed
\autocite{wesolowski2015}. A delay between player contribution and
winner identification by the authority is imposed, so that estimations
would be available only after contributions are no longer allowed.
Players or any other third party can engage before a deadline in the
collection of arbitrary data, e.g.~using social networks like Twitter,
to generate a seed, an essentially random bit string. Right after the
deadline, the authority publishes a commitment on an additional,
secretly chosen seed. Both seeds provide the input for a \emph{proof of
work}. A proof of work is computationally expensive to generate and thus
time-consuming. A delay is inevitable. However, due to its asymmetry,
the proof allows for efficient verification. Once the proof is found,
the winners are derived from it. The additional seed prevents dishonest
players to predict the results for different potential last-minute
contributions. One has to assume that the last honest contribution is
made sufficiently late to prevent the same attack from the authority.

Chow's online lottery protocol~\autocite{chow2005} published prior
to~\autocite{wesolowski2015} relies on a technique called \emph{delaying
function}~\autocite{goldschlag1998} that is similar to a proof of work,
but is not asymmetric and does not provide efficient verification. The
authority commits here on the concatenation of the players' commitments
on their secretly chosen number and derives then the winners. Players
can claim the reward by publishing the input data of their commitment.
Similar to~\autocite{wesolowski2015}, a late honest player commitment is
assumed to prevent a prediction by the authority. Then, all security
measures from the introduction are provided. The protocol requires
players to process the commitment of all other players and recompute the
delaying function in order to verify the random process, which is
impractical for large-scale lotteries.

Solutions for a scalable probabilistic verification of online
lotteries~\autocite{liu2007} or online voting~\autocite{dossogne2010}
have been presented based on a concatenation/aggregation over a tree
structure. In order to verify the result at the root of the tree,
players or voters can repeat the computation of intermediate results for
a predefined or random subset of all tree nodes. With increasing number
of verified nodes, the probability of a manipulated result at the root
node diminishes.

Other online lottery protocols introduce mutually distrusting,
non-colluding authorities to allow for a \emph{separation of powers}.
In~\autocite{kuacharoen2012}, a distinct auditor ensures secrecy and
immutability of the player's tickets and prevents the lottery authority
from adding illegitimately tickets. For this, blind signatures and
public-key encryption are employed. The protocol does not cover the
random process and its verification. Authorities are assumed not to
collude.

In~\autocite{fouque2001}, the secrecy of online lottery and voting
protocols is addressed at the same time. A mechanism based on
homomorphic encryption, distributed key generation and threshold
decryption is proposed. A set of mutually distrusting authorities have
to cooperate to decrypt the result of the random process or the voting.
A colluding set of dishonest authorities below the threshold cannot
reveal prematurely the result, i.e.~to add a winning ticket in the
lottery case. Players or voters are entitled to trust that the set of
dishonest, colluding authorities does not meet the threshold. Ideally,
the power to decrypt would be shared among all players or voters.
Practically, this is often not feasible due to scalability issues.

The \emph{Scalable and Secure Aggregation} (SPP) online voting
protocol~\autocite{gambs2011} builds also on distributed decryption and
employs a tree overlay network to improve the scalability. A small set
of authorities is randomly chosen among all voters. If too many of those
chosen voters are absent after the aggregation, the decryption threshold
cannot be reached and, consequently, a protocol termination is
impossible.

The potential of the Bitcoin blockchain~\autocite{bitcoin08} for a
distributed random process has been examined. However, it has been shown
that the manipulation of presumably random bits is realistic even with
limited computational capacity and financial
resources~\autocite{wesolowski2016}. An integration of the proof of work
from~\autocite{wesolowski2015} and an alternative crypto-currency
Ethereum~\autocite{wood2014} has been
proposed\footnote{\url{http://www.quanta.im}, \url{https://kiboplatform.net} (accessed 02/02/2017)}
with no practical solution yet for a verification due to the limitation
imposed by the blockchain.

\section{Preliminaries}\label{sec:preliminaries}

The starting point for the proposed protocol is the centralised online
lottery protocol of \textcite{chow2005}, recalled hereafter with an
alternative verification based on hash trees~\autocite{liu2007}. For the
proposed lottery protocol, we choose to distribute the random process to
all players. The overlay network comprising all players is provided by
the distributed hash table (DHT) Kademlia~\autocite{maymounkov2002} that
is described in Section~\ref{sec:kademlia}. The integration of these
building blocks is shown in Section~\ref{sec:distributed-lottery}.

\subsection{Centralised Online Lottery
Protocol}\label{sec:centralised-lottery}

The following presentation of Chow's
protocol~\autocites{chow2005}{liu2007} is reduced to aspects required
for our proposition. We use the following notation:

\begin{longtable}[]{@{}rl@{}}
\toprule
\(A\) & authority (Dealer in \autocite{chow2005})\tabularnewline
\(P_i\) & player, \(i\)-th out of \(n\)\tabularnewline
\(n_i\) & number in the set \(\mathbb{L}\) chosen by player
\(P_i\)\tabularnewline
\(r_i\) & random bit string of given length chosen by player
\(P_i\)\tabularnewline
\(\eta(\cdot)\) & cryptographic hash function, e.g.~SHA-3\tabularnewline
\(\eta_0(\cdot)\) & cryptographic hash function mapping any \(r_i\) to
\(\mathbb{L}\)\tabularnewline
\(\sigma_A(\cdot)\) & authority's signature scheme using key-pair
\((pk_A,sk_A)\)\tabularnewline
\bottomrule
\end{longtable}

\noindent
Chow's protocol implements a lottery in which every player~\(P_i\) has
to choose a number~\(n_i \in \mathbb{L}\) and send a commitment on it to
the authority~\(A\). \(A\) aggregates all commitments to a value \(h\).
That means, every~\(P_i\) contributes to~\(h\). The aggregate~\(h\) is
used as an input parameter for a \emph{delaying function} (DF)
preventing~\(A\) from early result estimations. The outcome of DF is
used to compute the winning number \(n_R\) with a \emph{verifiable
random function} and the secret key of~\(A\). Players do not have the
secret key required to compute~\(n_R\), but can verify~\(n_R\) using the
public key of~\(A\).

During the \emph{ticket purchase} phase, \(P_i\) acquires from \(A\) a
personal sequence number \(s_i\). \(P_i\) has to choose its number
\(n_i\) and a random bit string \(r_i\) to compute its commitment
\(\text{ticket}_i\) with bit string concatenation~\(||\) and XOR
operation~\(\veebar\). \(P_i\)~sends \(\text{ticket}_i\) to~\(A\) and
receives in return the signature \(\sigma_A(\text{ticket}_i)\) as a
receipt.

\[ \text{ticket}_i = s_i||(n_i \veebar \eta_0(r_i))||\eta(n_i||s_i||r_i) \]

The DF cannot be evaluated before \(h\) depending on all commitments is
given, which is ideally only after the purchase phase.
In~\autocite{chow2005}, the DF input parameter \(h\) is recursively
computed from all \(n\)~commitments with \(h = \eta(\text{chain}_n)\)
and \(\text{chain}_i = \eta(\text{chain}_{i-1}||\text{ticket}_i)\) with
an empty initial chain \(\text{chain}_0\). An alternative introduced
in~\autocite{liu2007} consists of a computation of~\(h\) using a
\(T\)-ary Merkle tree~\autocite{merkle1988} with \(\text{ticket}_i\)
assigned to the leaf tree nodes. In both cases, all \(\text{ticket}_i\)
are published to allow the verification of \(h\) by the players
requiring memory and computational resources of respectively \landau{n}
and \landau{\log_T(n)}.

Once the authority has published the verifiable winning number \(n_R\),
the \emph{reward claiming} phase begins in which players \(P_i\) with
\(n_i = n_R\) provide their sequence number~\(s_i\) and their secret
random value \(r_i\) to \(A\) via a secure channel. Upon verification of
the commitment \(\text{ticket}_i\) by \(A\), the reward is granted.
\(P_j\) with \(n_j \neq n_R\) may verify that their commitment
\(\text{ticket}_j\) has been used to compute \(h\) and are assumed to
have trust in the infeasibility of \(A\) to compute DF more than once
between the latest honest ticket contribution and the publication
of~\(n_R\).

\subsection{Distributed Hash Table Kademlia}\label{sec:kademlia}

The distributed hash table (DHT) Kademlia~\autocite{maymounkov2002}
provides efficient discovery of lottery players and routing which is a
precondition for the aggregation protocol in Section~\ref{sec:advokat}.
Therefore, a binary overlay network is established in which each player
\(P_i\) is assigned to a leaf node \(x_i\), that is a bit string of size
\(B\). The notation is as follows:

\begin{longtable}[]{@{}rl@{}}
\toprule
\(x\) & a Kademlia leaf node ID (KID) of size \(B\)\tabularnewline
\(B\) & size of a KID in bits, e.g.~160\tabularnewline
\(x_i\) & KID of player \(P_i\)\tabularnewline
\(d\) & node depth, i.e.~number of edges from the node to the tree
root\tabularnewline
\(\cSs(x,d)\) & subtree whose root is at depth \(d\) which contains leaf
node \(x\)\tabularnewline
\(\cS(x,d)\) & \emph{sibling subtree} of which the root is the sibling
of the root of \(\cSs(x,d)\)\tabularnewline
\(k\) & maximum number of contacts per Kademlia subtree\tabularnewline
\bottomrule
\end{longtable}

\noindent
The leaf node identifiers \(x \in \{0,1\}^B\) (\(B\) bits) span the
Kademlia binary tree of height \(B\) and are denoted KID. Each player
\(P_i\) joins the Kademlia overlay network using its KID defined as
\(x_i = \eta(t_i)\) with an authorization token \(t_i\) and the hashing
function \(\eta(\cdot)\). The value \(t_i\) is generated as part of the
ticket purchase. \(B\) is chosen sufficiently large, so that hash
collisions leading to identical KIDs for distinct players are very
unlikely. Consequently, the occupation of the binary tree is very
sparse.

A node in the tree is identified by its depth \(d \in \{0,\ldots,B\}\)
and any of its descendant leaf nodes with KID \(x\). A \emph{subtree}
\(\cSs(x,d)\) is identified by the depth \(d\) of its root node and any
of its leaf nodes \(x\). We overload the subtree notation to designate
as well the set of players assigned to leaves of the corresponding
subtree. Further, we introduce \(\cS(x,d)\) for the sibling subtree of
\(\cSs(x,d)\), so that \(\cSs(x,d) = \cSs(x,d+1)\cup\cS(x,d+1)\). The
entire tree is denoted~\(\cSs(x,0)\). We observe that
\(\forall d: P_i \in \cSs(x_i,d)\) and
\(\forall d: P_i \notin \cS(x_i,d)\).

Kademlia defines the distance \(d(x_i,x_j)\) between two KIDs as their
bit-wise XOR interpreted as an integer. In general, a player \(P_i\)
with KID \(x_i\) stores information on players with \(x_j\) that are
close to \(x_i\), i.e.~for small \(d(x_i,x_j)\). For this purpose,
\(P_i\) disposes of a set denoted \(k\)-bucket of at most \(k\) players
\(P_j \in \cS(x_i,d_j)\) for some
\(d_j > 0\).\footnote{Note that originally \autocite{maymounkov2002} the common prefix length $b$ is used to index $k$-buckets/sibling subtrees while we use the depth $d = b+1$ of the root of the subtree.}
See Fig.~\ref{fig:tree} for an example. The size of subtrees decreases
exponentially for growing depth \(d\). Hence, the density of known
players of corresponding \(k\)-buckets grows exponentially.

Kademlia ensures that the routing table, that is the set of all
\(k\)-buckets, is populated by player lookup requests for random KIDs to
the closest already known players. Requests are responded with a set of
closest, known players from the routing table. One lookup might require
multiple, consecutive request-response cycles. Further, Kademlia
provides requests to lookup and store values. All operations scale with
\landau{\log n}~\autocite{cai2013}. Kademlia is used by many BitTorrent
clients and as such well tested.

\tikzset{
  sarrow/.style    = {->, >=open triangle 90},
  darrow/.style    = {->, dashed,>=open triangle 90},
  point/.style     = {circle,draw,inner sep=0pt,minimum size=0pt},
  bigpoint/.style  = {fill=black,circle,draw,inner sep=0pt,minimum size=4pt},
  empty/.style     = {thick,gray,fill=black!25,circle,draw,minimum size=10pt},
  idleaf/.style    = {circle,draw,inner sep=1pt,minimum size=0.6cm,label={[yshift=0.8mm]below:$\cSs(x_i{,}3)$}},
  treeedge/.style  = {
    draw, 
    edge from parent path={(\tikzparentnode) -- (\tikzchildnode)}
  },
  mixp/.style      = {draw,rectangle,rounded corners}
  upbrace/.style   = {decorate,decoration={brace,mirror},thick},
  downbrace/.style = {decorate,decoration={brace},thick},
  highlightpath/.style = {very thick,draw=blue}
}

\begin{figure}[tb]
  \centering
  \begin{tikzpicture}
      [
          level 1/.style={sibling distance=40mm},
          level 2/.style={sibling distance=25mm},
          level 3/.style={sibling distance=13mm},
          edge from parent/.style={thick,draw=black!70},
          level distance=10mm,
      ]
      \node[point] (root) {}
      [
          every child node/.style={point}
      ]
          child {
              node (s01) {}
              child {
                  node (s02) {}
                  child {
                      node[empty] (s03)  {}
                      edge from parent
                      node[left] {0}
                  }
                  child[missing]
                  edge from parent
                  node[left] {0}
              }
              child {
                  node (s04) {}
                  child {
                      node[empty] (s05)  {}
                      edge from parent
                      node[left] {0}
                  }
                  child[missing]
                  edge from parent
                  node[right] {1}
              }
              edge from parent
              node[left=5pt] {0}
          }
          child {
              node {}
              child {
                  node {}
                  child {
                      node[idleaf] (x) {$x_i$}
                      edge from parent[highlightpath]
                      node[left] {0}
                  }
                  child {
                      node[empty] (s21) {}
                      edge from parent
                      node[right] {1}
                  }
                  edge from parent[highlightpath]
                  node[left] {0}
              }
              child {
                  node (s11) {}
                  child {
                      node[empty] (s12) {}
                      edge from parent
                      node[left] {0}
                  }
                  child {
                      node[empty] (s13) {}
                      edge from parent
                      node[right] {1}
                  }
                  edge from parent
                  node[right] {1}
              }
              edge from parent[highlightpath]
              node[right=5pt] {1}
          };
          \node[draw,dashed,fit=(s01) (s02) (s03) (s04) (s05),label=below:$\cS(x_i{,}1)$] (d3) {};
          \node[draw,dashed,fit=(s11) (s12) (s13),label=below:$\cS(x_i{,}2)$] (d2) {};
          \node[draw,dashed,fit=(s21),label=below:$\cS(x_i{,}3)$] (d1) {};
          \node[rectangle split, rectangle split parts=3, draw,rounded
          corners
          ]
          at ($(x)+(0,-1.8)$)
          (bucket)
          {
              $k$-bucket for $d=3$
              \nodepart{second}
              $k$-bucket for $d=2$
              \nodepart{third}
              $k$-bucket for $d=1$
          };
          \draw[darrow] (d1.east) -- ++(0.4,0) |- (bucket.one east);
          \draw[darrow] (d2.east) -- ++(0.2,0) |- (bucket.second east);
          \draw[darrow] (d3.west) -- ++(-0.2,0) |- (bucket.third west);
  \end{tikzpicture}
  \caption{Example of Kademlia $k$-buckets for KID $x_i = 100$ assuming $B = 3$. The sparse tree is partitioned into subtrees $\cS(x_i,d)$ with their root node depth $d$. The $k$-buckets for each $d$ contain at most $k$ players $P_j \in \cS(x_i,d)$.}
  \label{fig:tree}
\end{figure}
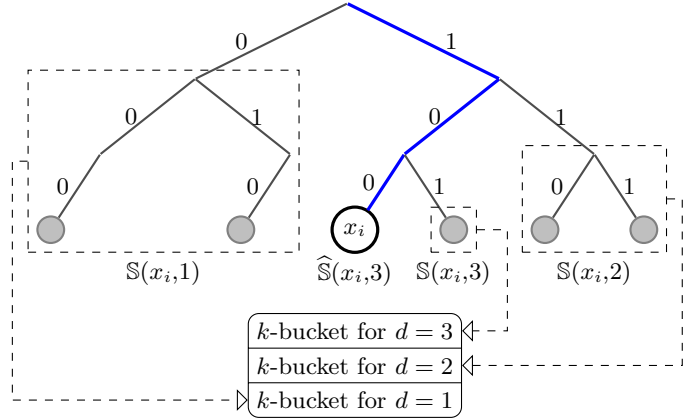

\section{Distributed Lottery}\label{sec:distributed-lottery}

We introduce now the lottery protocol. It is run by an authority that
handles the ticket purchase and carries out the distribution of the
reward upon winner verification, but not the random process itself. The
random process is distributed to all players using the protocol
described below. The description of the lottery protocol is given in
Section~\ref{sec:lottery-protocol}.

\subsection{Distributed Aggregation Protocol}\label{sec:advokat}

We present a distributed aggregation protocol based on Kademlia. It
relies on \textsc{Advokat} \autocite{riemann2017}, whose aggregation
algebra, distributed aggregation algorithm, and measures to increase its
Byzantine fault tolerance are briefly recalled.

\subsubsection{Aggregation Algebra}\label{aggregation-algebra}

\emph{Aggregates} are values to be aggregated, whether \emph{initial
aggregates}, constituting inputs from players, or \emph{intermediate
aggregates} obtained during the computation. The aggregation operation,
\(\oplus\), combines two \emph{child aggregates} to a \emph{parent
aggregate} in \(\mathbb{A}\), the set of aggregates. We assume
\(\oplus\) to be commutative. For the lottery, \(\oplus\) maps pairwise
bit strings provided by all players to one final bit string used to
determine the winners. The algebra is sufficiently flexible to cover a
broad range of aggregation-based applications and has been devised
initially for distributed online voting \autocite{riemann2017}.

Aggregates are manipulated through \emph{aggregate containers}, i.e.~a
data structure that contains next to the aggregate itself the context of
the ongoing computation. The aggregate container of an aggregate \(a\)
associates \(a\) to a Kademlia subtree \(\cSs(x,d)\) and ensures
integrity and verifiability of the aggregation. It has the following
attributes:

\begin{longtable}[]{@{}rl@{}}
\toprule
\(h\) & hash \(\eta(\cdot)\) of the entire aggregate container, but
\(h\)\tabularnewline
\(a\) & aggregate, \(a = a_1 \oplus a_2\)\tabularnewline
\(c\) & counter of initial aggregates in \(a\),
\(c = c_1 + c_2\)\tabularnewline
\(c_1\), \(c_2\) & counter of initial aggregates of child
aggregates\tabularnewline
\(h_1\), \(h_2\) & container hashes of child aggregates\tabularnewline
\(\cSs(x,d)\) & identifier of subtree whose initial aggregates are
aggregated in \(a\)\tabularnewline
\bottomrule
\end{longtable}

\noindent
Similar to the aggregation of aggregates, one or two aggregate
containers of \(a_1\), \(a_2\) can be aggregated to a parent aggregate
container. To inherit the commutativity of the aggregation of aggregates
\(\oplus\), \((h_1,c_1)\) and \((h_2,c_2\)) must be sorted in
e.g.~ascending order of the child hashes \(h_1\) and \(h_2\).

\subsubsection{Distributed Aggregation
Algorithm}\label{sec:basic-aggregation}

Using the aggregation operator \(\oplus\), every player \(P_i\) computes
the intermediate aggregate for all the parent nodes from its
corresponding leaf node \(x_i\) up to the root node of the Kademlia
overlay tree. The aggregates used to compute any intermediate aggregate
of a given subtree \(\cSs(x_i,d)\) are given by its child nodes'
aggregates of \(\cSs(x_i,d+1)\) and \(\cS(x_i,d+1)\). Hence, aggregates
have to be exchanged between players of the sibling subtrees and
Kademlia's \(k\)-buckets provide the required contact information.

The aggregation is carried out in \(B\) epochs, one tree level at a
time. Epochs are loosely synchronized, because players may have to wait
for intermediate aggregates to be computed in order to continue. First,
every player \(P_i\) computes a container for its initial aggregate
\(a_i\). The container is assigned to represent the subtree
\(\cSs(x_i,B)\) with only \(P_i\). In each epoch for \(d=B,\ldots,1\),
every player \(P_i\) requests from a random \(P_j \in \cS(x_i,d)\) the
aggregate container of subtree \(\cS(x_i,d)\). With the received
container of \(\cS(x_i,d)\) and the previously obtained of
\(\cSs(x_i,d)\), player \(P_i\) computes the parent aggregate container,
that is then assigned to the parent subtree \(\cSs(x_i,d-1)\). If
\(\cS(x,d) = \emptyset\) for any \(d\), the container of \(\cSs(x,d-1)\)
is computed only with the aggregate container of \(\cSs(x,d)\) from the
previous epoch.

After \(B\) consecutive epochs, player \(P_i\) has computed the root
aggregate \(a_R\) of the entire tree \(\cSs(x_i,0)\) that contains the
initial aggregates of all players. If all players are honest, the root
aggregate is complete and correct. Due to the commutativity of the
container computation, all players find the same hash \(h_R\) for the
container of the root aggregate \(a_R\). An individual verification is
implicitly given, because every player computes \(a_R\) starting with
its \(a_i\).

\subsubsection{Byzantine Fault-Tolerance}\label{sec:robust-aggregation}

\newcommand*\circled[1]{\tikz[baseline=(char.base)]{
            \node[shape=circle,draw,inner sep=1pt,font=\tiny] (char) {#1};}}
\newcommand*\circledtext[1]{\tikz[baseline=(char.base)]{
            \node[shape=circle,draw,inner sep=1pt,font=\scriptsize] (char) {#1};}}

The distributed aggregation is very vulnerable to aggregate corruptions
leading to erroneous root aggregates containers. We present intermediate
results to safeguard the aggregation. Please refer to
\autocite{riemann2017} for a more in-depth discussion. For the attack
model, we assume a minority of dishonest (Byzantine) players controlled
by one adversary that aims to interrupt the aggregation, and manipulate
root aggregates. Dishonest players can behave arbitrarily. Like in
Kademlia, time-outs are used to counter unresponsive players.

To prevent Sybil attacks, it must be ensured that a player a) cannot
choose on its own discretion its tree position given by the leaf
node~\(x_i\) but b) can proof its attribution to~\(x_i\)
\autocite{baumgart2007}. Every player~\(P_i\) generates a key
pair~\((pk_i,sk_i)\) which must be signed by~\(A\). Hence, \(P_i\)
sends~\(pk_i\) to \(A\) during the ticket purchase and receives the
signature of \(A\) to be used as the authorization token
\(t_i = \sigma_A(pk_i)\). The KID \(x_i = \eta(t_i)\) is derived from
\(t_i\) and is neither chosen unilaterally by~\(A\) nor by~\(P_i\).
Eventually, players provide for every message \(m\) exchanged among
players the signature of the sender~\(\sigma_i(m)\), its public key
\(pk_i\) to verify~\(\sigma_i(m)\), and the authorization token~\(t_i\)
to verify~\(pk_i\).

Moreover, a dishonest authority shall be prevented to add new players
after the aggregation has started and dishonest players to delay their
contributions after predefined, global deadlines. In order to suppress
both, signatures of those players are considered invalid, who are at the
start of the aggregation not in the corresponding \(k\)-bucket even
though the bucket contains less than \(k\) players and should be
exhaustive.

Further, player signatures are employed to detect deviations from the
protocol. For every computed aggregate container of \(\cSs(x_i,d)\) with
hash \(h\) and counter~\(c\), player \(P_i\) produces an aggregate
container signature \(\sigma_i(h,d,c)\). Other players can verify the
signature using \(pk_i\) and verify using \(x_i = \eta(t_i)\) that
\(P_i \in \cSs(x_i,d)\). Hence, \(pk_i\) and~\(t_i\) must be provided
along every signature~\(\sigma_i(h,d,c)\).

\tikzset{
  xleaf/.style    = {circle,draw,inner sep=1pt,minimum size=0.6cm},
}

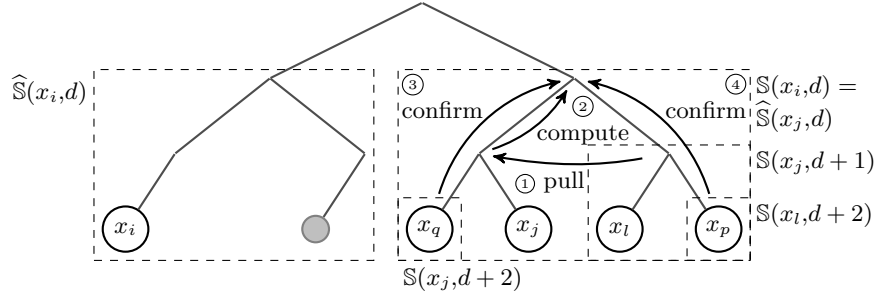
\begin{figure}
  \centering
  \begin{tikzpicture}
      [
          level 1/.style={sibling distance=40mm},
          level 2/.style={sibling distance=25mm},
          level 3/.style={sibling distance=13mm},
          edge from parent/.style={thick,draw=black!70},
          level distance=10mm,
          lfs/.style={font=\footnotesize},
      ]
      \node[point] (root) {}
      [
          every child node/.style={point}
      ]
          child {
              node (s01) {}
              child {
                  node (s02) {}
                  child {
                      node[xleaf] (s03)  {$x_i$}
                  }
                  child[missing]
              }
              child {
                  node (s04) {}
                  child {
                      node[empty] (s05)  {}
                  }
                  child[missing]
              }
          }
          child {
              node (sm1) {}
              child {
                  node (sm2) {}
                  child {
                      node[xleaf] (x) {$x_q$}
                      edge from parent
                      node[left] (smc2) {}
                  }
                  child {
                      node[xleaf] (s21) {$x_j$}
                  }
              }
              child {
                  node (s11) {}
                  child {
                      node[xleaf] (s12) {$x_l$}
                  }
                  child {
                      node[xleaf] (s13) {$x_p$}
                      edge from parent
                      node[right] (sm3) {}
                  }
              }
          };
          \node[draw,dashed,fit=(s01) (s03) (s04)] (bb) {};
          \node[below left, inner sep=2pt] at (bb.north west) {$\cSs(x_i{,}d)$};
          \node[draw,dashed,fit=(sm1) (x) (s13)] (b0) {};
          \node[below right, inner sep=2pt,align=left] at (b0.north east) {$\cS(x_i{,}d)=$\\$\cSs(x_j{,}d)$};
          \node[draw,dashed,fit=(s11) (s12) (s13)] (b1) {};
          \node[below right, inner sep=2pt] at (b1.north east) {$\cS(x_j{,}d+1)$};
          \node[draw,dashed,fit=(s13)] (b2) {};
          \node[below right, inner sep=2pt] at (b2.north east) {$\cS(x_l{,}d+2)$};
          \node[draw,dashed,fit=(x)] (b22) {};
          \node[below right, inner sep=2pt] at (b22.south west) {$\cS(x_j{,}d+2)$};
          \path (s11) edge[bend left=10,thick,->,>=stealth',shorten >=5pt, shorten <=10pt] node[label={[lfs,shift={(-9pt,4pt)}]below:\circled{1} pull}] {} (sm2);
          \path (sm2) edge[bend right=20,thick,->,>=stealth',shorten >=5pt, shorten <=5pt] node[label={[lfs,align=center,shift={(-6pt,2pt)}]right:\circled{2}\\compute}] {} (sm1);
          \path (s13) edge[bend right=30,thick,->,>=stealth',shorten >=5pt, shorten <=5pt] node[label={[lfs,align=right,shift={(-6pt,10pt)}]right:\circled{4}\\confirm}] {} (sm1);
          \path (x) edge[bend left=30,thick,->,>=stealth',shorten >=5pt, shorten <=5pt] node[label={[lfs,align=left,shift={(6pt,10pt)}]left:\circled{3}\\confirm}] {} (sm1);
  \end{tikzpicture}
  \caption{$P_j$ with $x_j$ produces a confirmed aggregate container of $\cS(x_i,b)$. This scheme applies to all tree levels with possibly large subtrees to request from. If the subtrees $\cS(x_j,d+2),\cS(x_l,d+2)$ are empty, the depth is further increased until a non-empty subtree may be found.}
  \label{fig:robust-pull}
\end{figure}

The impact of dishonest players is limited by redundant requests to
confirm a computed so-called \emph{candidate container} using signatures
of other players on the same container hash as depicted in
Fig.~\ref{fig:robust-pull}. Next to the proper signature
(\circledtext{2}) on \(h\) and \(h_1\), a signature on \(h\) from a
player in each child subtree (\circledtext{3} and \circledtext{4}) and
one on \(h_2\) (\circledtext{1}) must be provided for a confirmation if
the respective subtree is non-empty which can be determined using
Kademlia lookup requests. The number of distinct signatures on \(h\),
here \(3\), is a security parameter denoted \(l\).

Only the \emph{confirmed container} including the signatures is used to
respond to requests from players in the sibling subtree. If a
confirmation is not possible, e.g.~due to non-cooperating dishonest
players, the confirmed child containers are provided instead, so that
the receiver can compute the aggregation on its own.

If confirmation requests reveal diverging containers, a majority vote
using the number of distinct signatures for every container hash is
used. If another container than the previously computed is selected and
if \(h_2\) differs, then the request for the sibling aggregate container
(\circledtext{1}) is repeated, otherwise, the previous epoch is repeated
allowing for a recursive correction.

The majority vote confirms for subtrees with many players with great
probability the aggregate container of the honest players. The
attribution of KIDs \(x_i\) to players \(P_i\) is random, so that a
global minority of dishonest players is uniformly distributed over all
subtrees and a honest majority can be assumed for most local subtrees.

Though, dishonest players may have a local majority in subtrees with
only few players. Here, an analysis of the signatures of confirmed
containers allows honest players to detect dishonest behaviour in the
following cases with certainty. Given two signatures \(\sigma_i(h,d,c)\)
and \(\sigma_i(h',d,c)\) from the same player~\(P_i\) with different
hashes~\(h\neq h'\), \(P_i\) deviated from the protocol with certainty
if \(c \le l\). Either \(P_i\) signed two distinct initial aggregate
containers or accepted a non-confirmed container. For \(c > l\), there
is a non-zero probability that \(P_i\) is honest, but may have been
tricked. A manipulation may not be detected or only later during the
recursive correction. The number of distinct signatures \(l\) can be
increased to detect manipulations with certainty for higher \(c\), and
may depend on the player configuration in the respective subtree.

At last, the root aggregate container~\(a_R\) shall be confirmed more
often, i.e.~more signatures on its hash \(h\) are gathered from
different players, to increase the confidence that it has been adopted
by the majority of honest players.

\subsection{Lottery Protocol}\label{sec:lottery-protocol}

The proposed protocol allows for a lottery with playing mode CL or
LO~\autocite{kuacharoen2012}:

\setlist[description]{style=nextline}

\begin{description}
\tightlist
\item[Classic Lottery (CL)]
Rewards are distributed with respect to a randomly ordered list of all
players.
\item[Lotto (LO)]
Rewards are distributed based on the secret, prior choice of each
player.
\end{description}

\noindent
The protocol has six phases of which \emph{ticket purchase},
\emph{reward claiming} and \emph{winner verification} follow closely
Chow's protocol~\autocite{chow2005}. Its model provides for an
authority~\(A\), a tracker and players~\(P_i\). For CL,
\(\eta_0(\cdot)\) is identical to~\(\eta(\cdot)\), so that the root
aggregate \(a_R\) of the distributed aggregation is in the domain of the
KIDs~\(x\).

\subsubsection{Setup}\label{setup}

\begin{enumerate}
\def\labelenumi{\arabic{enumi}.}
\tightlist
\item
  \(A\) generates a key-pair \((pk_A,sk_A)\) and chooses a random bit
  string \(r_A\).
\item
  \(A\) publishes the ticket purchase deadline, \(pk_A\),
  \(\eta(\cdot)\), \(\eta_0(\cdot)\) and \(\eta(\eta(r_A))\). Further,
  \(A\) specifies the duration of the aggregation epoch for each tree
  level.
\end{enumerate}

\subsubsection{Ticket Purchase}\label{ticket-purchase}

\begin{enumerate}
\def\labelenumi{\arabic{enumi}.}
\tightlist
\item
  \(P_i\) picks a random string \(r_i\), and for CL its number \(n_i\).
  It generates \((pk_i,sk_i)\).
\item
  \(P_i\) sends \(pk_i\) to the authority and obtains in return a
  sequence number~\(s_i\) and its authorization token
  \(t_i = \sigma_A(pk_i)\).
\item
  \(P_i\) computes \(x_i = \eta(t_i)\) and connects to the Kademlia DHT
  using an already connected contact provided by the authority or a
  separate tracker.
\item
  \(P_i\) prepares its initial aggregate
  \(a_i = \eta(\text{ticket}_i)\). For CL,
  \(\text{ticket}_i = s_i||r_i\) and for LO,
  \(\text{ticket}_i = s_i||(n_i\veebar\eta_0(r_i))||\eta(n_i||s_i||r_i)\),
  c.f.~Section~\ref{sec:centralised-lottery}.
\end{enumerate}

\subsubsection{Distributed Random
Process}\label{distributed-random-process}

\begin{enumerate}
\def\labelenumi{\arabic{enumi}.}
\tightlist
\item
  After the ticket purchase deadline, \(A\) publishes the number of sold
  tickets \(n\).
\item
  All \(P_i\) compute jointly the root aggregate \(a_R\). The
  \(\oplus\)-operation is given by \(a_i \oplus a_j = \eta(a_{ij})\)
  with \(a_{ij} = a_i || a_j\), if~\(a_i < a_j\),
  otherwise~\(a_{ij} = a_j || a_i\). It is a commutative variant of the
  binary Merkle tree scheme proposed in~\autocite{liu2007}.
\item
  Proofs of protocol deviation in form of pairs of signatures are sent
  to \(A\) that can reveal the corresponding players and revoke their
  right to claim a reward.
\end{enumerate}

\subsubsection{Winner Identification}\label{winner-identification}

\begin{enumerate}
\def\labelenumi{\arabic{enumi}.}
\tightlist
\item
  \(A\) requests the root aggregate of multiple random \(P_i\) until a
  considerably large majority of the sample confirmed one \(a_R\).
\item
  \(A\) publishes \(a_R\), \(r_A\) and the winning number
  \(n_w = \eta_0(a_R) \veebar \eta_0(r_A)\).
\item
  For CL, \(A\) computes all \(x_i = \eta(t_i)\), orders all \(P_i\) by
  their Kademlia XOR distance \(d(n_w,x_i) = n_w \veebar x_i\) and
  players on a par by~\(n_w \veebar s_i\), and publishes as many ordered
  \(x_i\) as there are rewards. For LO, winners \(P_i\) have
  \(n_i = n_w\).
\end{enumerate}

\subsubsection{Reward Claiming}\label{reward-claiming}

\begin{enumerate}
\def\labelenumi{\arabic{enumi}.}
\tightlist
\item
  The winner \(P_i\) sends all its confirmed aggregate containers to
  \(A\) to proof their participation. For LO, \(P_i\) must also provide
  its \(\text{ticket}_i\) and \((s_i,r_i)\).
\item
  Proofs are published for verification by other players.
\end{enumerate}

\subsubsection{Winner Verification}\label{winner-verification}

\begin{enumerate}
\def\labelenumi{\arabic{enumi}.}
\tightlist
\item
  \(\eta(\eta(r_A))\) is computed for comparison with the previously
  published value and \(n_w\) is verified.
\item
  Players verify that winner \(P_i\) participated in the aggregation by
  comparing its published containers with their computed containers.
\item
  For CL, player verify the order of the published winners and compare
  it to their own positioning. For LO, \(\text{ticket}_i\) is reproduced
  for the given \((s_i,r_i)\) and its hash must equal \(a_i\) found in
  the published confirmed aggregate containers.
\item
  If the rewards depend on the number of sold tickets \(n\), \(n\) is
  compared to the counter \(c\) of the root aggregate container.
\end{enumerate}

\section{Evaluation}\label{sec:evaluation}

We analyse the protocol with respect to the security properties
introduced in Section~\ref{sec:introduction} under the adversary model
from Section~\ref{sec:robust-aggregation} of an adversary \(D\)
controlling a fraction \(b < 0.5\) of dishonest players of \(n\) players
in total. The performance of the protocol depends upon~\(b\), the
security parameter \(l\) and the distribution of honest and dishonest
players over the tree. We assume that~\(D\) and~\(A\) collude.

\subsubsection{Most Likely Scenario}\label{most-likely-scenario}

Due to the uniform player distribution and for a reasonably sized \(b\),
\(D\) has most likely a dishonest majority only in subtrees with large
depth \(d > 1\) containing only a small number \(n'\) of players. \(l\)
can be adjusted to detect container manipulations of subtrees with
\(n' \le l\) using signatures. Most likely, all dishonest players have
to provide a container with their signature to at least one honest
player, which corresponds to a commitment to their \(\text{ticket}_i\),
before \(D\) can learn all containers for a given depth~\(d\).

\begin{enumerate}
\def\labelenumi{\arabic{enumi}.}
\tightlist
\item
  The \emph{correctness} of the random process and its implicit
  \emph{verification}~\autocite{liu2007} due to the distributed
  computation is with great probability ensured, because \(D\) cannot
  change or add tickets after a prediction becomes possible.
\item
  The \emph{privacy} of players is ensured. Other players cannot learn
  the identity of each other from the exchanged
  messages.\footnote{The leak of the identity due to the communication channel, e.g. by the IP address, may be solved using privacy networks like Tor and is out of the scope of this paper.}
\item
  The authorization token \(t_i\) ensures \emph{eligibility}. A
  participation after the aggregation has started even with a valid
  \(t_i\) is unlikely, because honest players close in the tree deny
  belated players and do not confirm their containers.
\item
  The commitment scheme for LO provides \emph{confidentiality}, because
  number~\(n_i\) of~\(P_i\) cannot be revealed without knowledge of the
  secret~\(r_i\)~\autocite{chow2005}.
\item
  The counter \(c\) of the root aggregate container allows to examine
  the \emph{completeness} of the reward.
\end{enumerate}

\subsubsection{Worst Case Scenario}\label{worst-case-scenario}

The distribution of honest and dishonest players is highly unbalanced.
We assume a majority of dishonest players in a subtree~\(\cSs(x_e,d)\)
for some~\(d\) with~\(n' > l\). Neither the majority nor the
confirmation criterion prevent a manipulation with certainty. The local
minority of honest players may be excluded from the aggregation unable
to proof their participation. As the manipulation is bounded locally,
correctness and eligibility are only locally violated.

If \(D\) has further in all other non-sibling subtrees~\(\cSs(\cdot,d)\)
at least one dishonest player to provide the local aggregate container,
\(D\) can compute with the secret~\(r_A\) from~\(A\) the winning
number~\(n_w\) while the container for \(\cSs(x_e,d)\) is not yet known
to honest players and may be altered to change \(n_w\). A proof of work
is required to choose a particular \(n_w\). The correctness is not
ensured.

The distribution of~\(n'\) honest or dishonest players to
\(\cSs(x_e,d)\) and its sibling subtree~\(\cS(x_e,d)\) follows the
Binomial distribution \(B(n',p)\) with \(p = 0.5\) and a variance of the
ratio of players in~\(\cSs(x_e,d)\) of~\(p^2/n'\). As a result, the
probability of a local dishonest majority decreases reciprocally
in~\(n'\). \(n'\) decreases for increasing~\(d\), but for large~\(d\),
it is unlikely to have a dishonest player in all non-sibling subtrees
for the limited number of dishonest players.

\subsubsection{Scalability}\label{scalability}

Kademlia's communication and memory resources
are~\landau{\log n}~\autocite{cai2013}. The same applies to the
distributed aggregation and its verification \autocite{riemann2017} if
upper bounds are defined for the number of attempts and stored container
candidates of the confirmation and correction mechanism of the
distributed aggregation.

\section{Conclusion}\label{sec:conclusion}

We have presented a novel online lottery protocol that relies on a
distributed random process carried out by all players in a peer-to-peer
manner. Players are assumed to participate throughout the random
process. Unlike Chow's protocol \autocite{chow2005}, it allows for both
classic lottery and lotto. It provides correctness and verification of
the random process based on the assumption of a well-distributed
minority of dishonest players. In the most likely scenario, the
correctness of the random process is based on an information theoretical
secure sharing scheme instead of assumptions on the communication or
computational capacities of the authority or the adversary. Further,
cryptography has been reduced to asymmetric encryption and signatures.
As in many distributed protocols~\autocites{bitcoin08}{gambs2011}, the
provided security is probabilistic, which may be acceptable for a
lottery. We leave for future work a quantitative analysis of the impact
of the adversary.

A basic demonstrator has been implemented to carry out a classical
lottery. The authority has been omitted in favour of free participation.
Redundant requests for Byzantine fault-tolerance are not covered yet.
Based on HTML5, it runs in the browser. The implementation of
\textsc{Adavokat} is based on the Kademlia library
\texttt{kad}\footnote{\url{http://kadtools.github.io/}, v1.6.2 released on November 29, 2016}
and was tested previously with up to 1000 simulated nodes
\autocite{riemann2017}. Message passing among players relies on WebRTC
allowing for browser-to-browser communication. Tests have been run with
few players at this stage.

\subsubsection{Acknowledgments}\label{acknowledgments}

The authors would like to thank Pascal Lafourcade and Matthieu Giraud
for fruitful discussions concerning the security of the lottery protocol
and the underlying distributed aggregation algorithm.

\section*{References} 
\printbibliography[heading=none]

\vfill
\end{document}